# Surface Modification and Subsequent Fermi Density Enhancement of Bi(111)

Kuanysh Zhussupbekov,* Killian Walshe, Brian Walls, Andrei Ionov, Sergei I. Bozhko, Andrei Ksenz, Rais N. Mozhchil, Ainur Zhussupbekova, Karsten Fleischer, Samuel Berman, Ivan Zhilyaev, David D. O'Regan,* and Igor V. Shvets*

**ABSTRACT:** Defects introduced to the surface of Bi(111) break the translational symmetry and modify the surface states locally. We present a theoretical and experimental study of the 2D defects on the surface of Bi(111) and the states that they induce. Bi crystals cleaved in ultrahigh vacuum (UHV) at low temperature (110 K) and the resulting ion-etched surface are investigated by low-energy electron diffraction (LEED), X-ray photoelectron spectroscopy, ultraviolet photoelectron spectroscopy (UPS), and scanning tunneling microscopy (STM) as well as spectroscopy (STS) techniques in combination with density functional theory (DFT) calculations. STS measurements of cleaved Bi(111) reveal that a commonly observed bilayer step edge has a lower density of states (DOS) around the Fermi level as compared to the atomic-flat terrace. Following ion bombardment, the Bi(111) surface reveals anomalous behavior at both 110 and 300 K: Surface periodicity is observed by LEED, and a significant increase in the number of bilayer step edges and energetically unfavorable monolayer steps is observed by STM. It is suggested that the newly exposed monolayer steps and the type A bilayer step edges result in an increase to the surface Fermi density as evidenced by UPS measurements and the Kohn−Sham DOS. These states appear to be thermodynamically stable under UHV conditions.

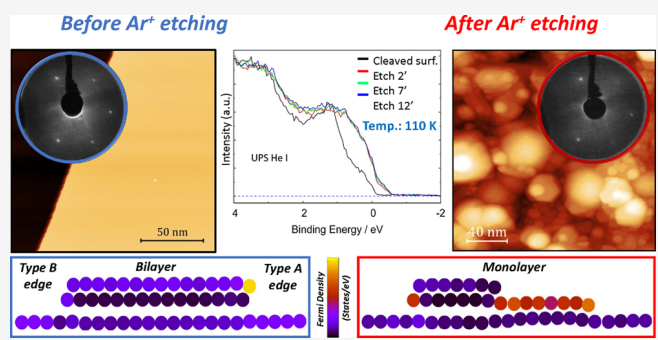

## ■ INTRODUCTION

In recent years, there has been increased attention on layered materials because of their wide range of applications in energy- and electronics-related fields.[1−6] Bismuth is one such material, attracting interest because of its topological properties.[7−11] Some Bi-based compounds are topological insulators and find thermal and catalytic applications.[12−17] The electronic structure[18−24] and formation of a charge-density wave (CDW) in both bulk Bi and ultrathin Bi films have previously been investigated by synchrotron radiation angle-resolved photoemission spectroscopy (ARPES).[25−27]

Bi has a rhombohedral crystal structure which can be described in terms of a small deformation of a simple cubic lattice.[28−30] The shifting of the (111) atomic planes leads to alternating spacing and chemical bonds; covalent bonds are present when layers are in close proximity to each other and van der Waals bonds when they are further apart, resulting in the formation of a layered structure. These deformations are energetically favorable due to a Peierls transition.[31−33] As a result of this Peierls transition, there is a gap in the electron dispersion in the vicinity of the Γ point of the Brillouin zone, while the spin−orbit interaction (SOI) plays a significant role in the formation of the electron spectrum. The energy of the SOI is comparable with the energy scales of the Peierls instability.[33] Bi has a lattice parameter of 4.54 Å; the interplanar distances in the [111] direction are $d_1$ = 1.59 Å and $d_2$ = 2.34 Å.[34,35] It has been shown in numerous experiments that Bi single crystals cleave along the (111) plane, breaking the van der Waals bonds between bilayers.[7,8,25,36]

One of the first studies of cleaved Bi(111) by scanning tunneling microscopy (STM) was conducted by Edelman et al.[36] They observed a new type of defect on Bi(111): twin interlayers in which a minimal width is observed. The width is governed by the matching of atomic planes at each side of the twin interlayer. Recently, bismuth and bismuth compounds have been widely studied for their topological properties.[9,25,37,38] Unlike some of its compounds (e.g., $Bi_2Se_3$ and $Bi_2Te_3$), bismuth itself is not a first-order topological insulator.[39] It therefore does not host 2D helical surface states. However, recent studies have revealed 1D topological edge states in a variety of different geometries at bismuth surfaces, including the (114) surface,[40] screw dislocations,[41] and type-A bilayer step edges at the (111) surface.[8] Bi(111) can have two types of bilayer step edges: armchair or zigzag (see Figure 6b). The zigzag edges terminate with an atom from the top of the bilayer (type A) while the armchair edges terminate with an atom from the bottom of the bilayer (type B). It has been shown by using STM and scanning tunneling spectroscopy (STS) that zigzag edges (type A) exhibit one-dimensional topological edge states and thus a higher density of states at the Fermi level.[7,8,10,21,42] These 1D topological states have been attributed to bismuth being a "higher order" topological material.[7] These discoveries highlight the importance of studying the electronic structure of 2D defects/1D structures such as step edges at bismuth surfaces.

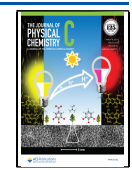

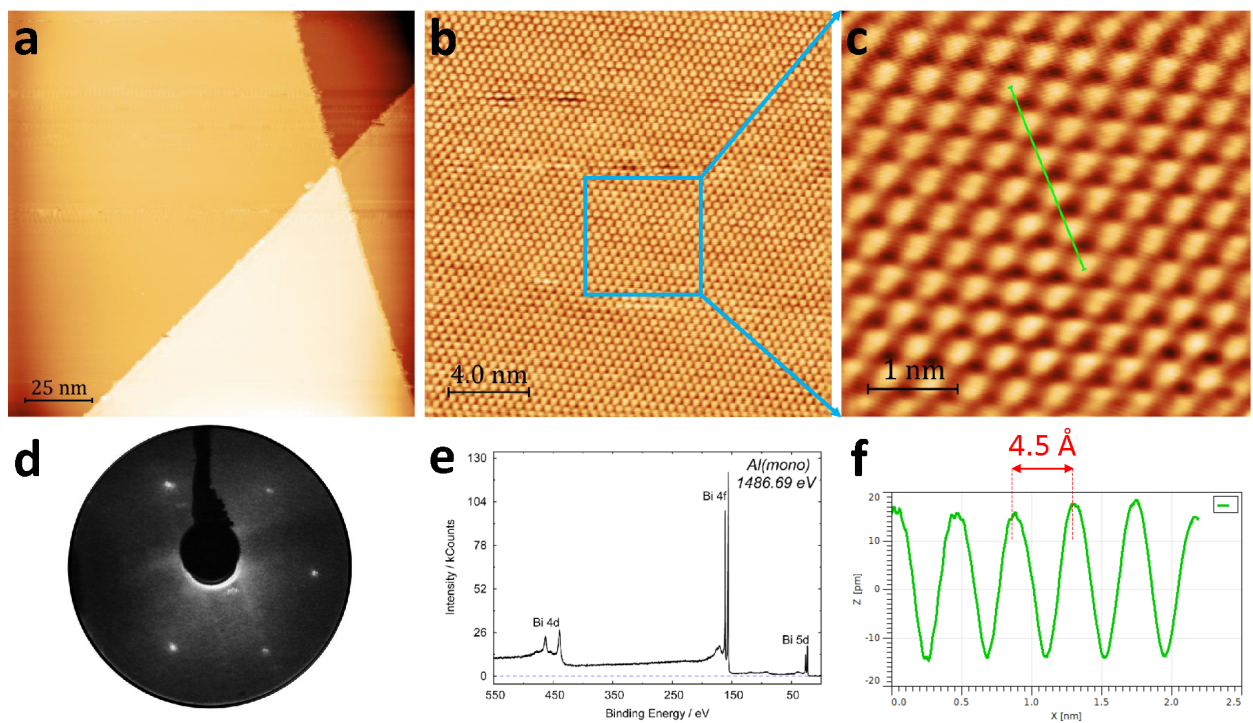

**Figure 1.** STM, LEED, and XPS of the cleaved Bi(111) surface. (a) Large-scale STM images of Bi(111) surface cleaved *in situ* at 110 K in UHV (150 × 150 nm$^2$, $V$ = 0.4 V and $I$ = 670 pA). Panels (b) and (c) illustrate the atomic resolution of the terrace in image (a). Scale and scanning parameters for (b) and (c) are 20 × 20 nm$^2$, $V$ = 2 V, $I$ = 80 pA and 4.5 × 4.5 nm$^2$, $V$ = 2 V, $I$ = 80 pA, respectively. (d) LEED image of the cleaved Bi(111) in UHV showing a high order of the surface. (e) XPS spectrum demonstrates that this surface is atomically clean. (f) Line profile of interatomic spacing of approximate periodicity 4.5 Å across the green line labeled in (c).

The primary motivation underpinning this study relates to the formation of defects on the surface, their evolution in time, and their effect on the electronic surface states.[43,44] Using ion bombardment, we can reduce the degree of order. This can result in anomalous monolayer planes and A-type bilayer steps coexisting with the B-type bilayer step, which dominate the pristine cleaved Bi(111) surface. In this work, we investigate and contrast STM, STS, low-energy electron diffraction (LEED), and photoemission measurements of a cleaved Bi(111) surface before and after ion etching. The effect of ion bombardment on the atomic and electronic surface structure is demonstrated, by using STM, to produce a surface with a higher density of monolayer steps and bilayer step edges. Ultraviolet photoelectron spectroscopy (UPS) measurements display an increase in intensity near the Fermi level by approximately a factor of 3. Density functional theory (DFT) calculations are qualitatively consistent with the UPS measurements, demonstrating an increase in the Kohn−Sham density of states at the Fermi level (Fermi density) for both the monolayer step and the type A bilayer step edge—which are shown to be induced by sputtering—compared to the type B bilayer steps.

■ EXPERIMENTAL AND COMPUTATIONAL SECTION

**Experimental Details.** Experimental measurements were performed across several ultrahigh-vacuum (UHV) systems. A vacuum suitcase with a base pressure of 2 × 10$^{-10}$ mbar was utilized for transferring crystals between STM and X-ray photoelectron spectroscopy (XPS) systems under UHV conditions. The STM used in this study is a commercial low-temperature system from Createc with a base pressure of 5 × 10$^{-11}$ mbar. All of the STM images were obtained at liquid nitrogen temperature (77 K) in constant-current mode (CCM). The preparation chamber of the UHV system is fitted with a cooling/heating stage, LEED, and an ion gun for sputtering. The STM tips used were of [001]-oriented single-crystalline tungsten, which electrochemically etched in NaOH.[45] The bias was applied to the sample with respect to the tip. XPS measurements were performed by using two XPS systems: an Omicrometer MultiProbe and a Kratos XPS. Both systems utilize monochromatic Al K$\alpha$ ($E$ = 1486.7 eV) with an instrumental resolution of 0.6 eV and a chamber base pressure of 5 × 10$^{-11}$ mbar. The valence spectra were measured by UPS with an excitation energy of He I (21.2 eV). The work function was determined by the cutoff of the secondary electrons. LEED measurements were taken across two systems: one for room temperature and the second for low-temperature measurements.

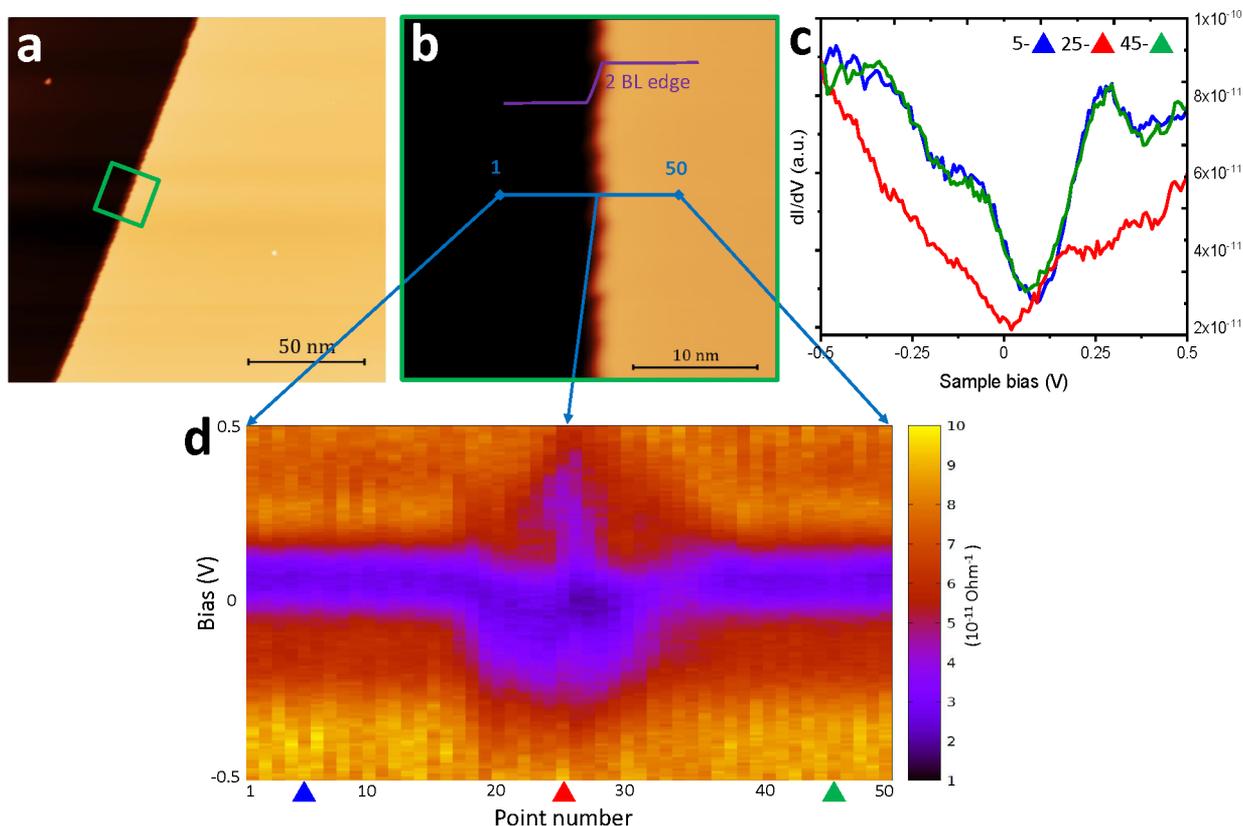

**Figure 2.** STS investigation of band profile across the double bilayer step of Bi(111). (a) Large-scale STM image (150 × 150 nm$^2$, $V$ = 1.2 V, and $I$ = 70 pA). (b) STM image of the green square labeled in (a) (30 × 30 nm$^2$, $V$ = 1.0 V, and $I$ = 80 pA). The blue line (15 nm) indicates where the line spectroscopy has been performed (stabilization parameters $V$ = 1.2 V and $I$ = 70 pA). (c) Representative d$I$/d$V$ spectra measured on and away from step edge. (d) 2D plot of tunneling spectra across the blue line (d$I$/d$V$ in bias range ±0.5 V) which demonstrates a suppression of the LDOS on the edge of the two bilayer steps.

Experimental measurements were conducted on ultrahigh-purity bismuth single crystals (residual resistance ratio (RRR): 800). The surface was prepared by cleaving the crystal in the load lock of the UHV system at a pressure of ∼5 × 10$^{-8}$ mbar at ∼110 K. The sample was promptly (within 5−10 s) transferred to the preparation chamber, where the pressure quickly returns to the base pressure of ∼2 × 10$^{-10}$ mbar. All stages of the sample preparation were monitored in situ by LEED and/or XPS. The surfaces with well-ordered LEED patterns demonstrating the (1 × 1) structure and/or without any contamination, concluded by XPS, were used in this study (see Figure 1d,e).

**Computational Details.** First-principles quantum-mechanical simulations of Bi steps and 2D defects were performed by using the PWscf DFT code of the Quantum Espresso suite.[46,47] Simulations were performed by using a slab geometry consisting of 10 layers (5 bilayers) with each layer containing 21 atoms, arranged linearly in the Bi[$\overline{1}$10] direction, with additional partial layers added to one surface of the slab to form the desired step edge structures. An out-of-plane vacuum spacing of 10 Å was used. The initial geometric parameters used in simulations were taken from experimental results found in the literature.[34] The atomic positions and cell dimensions were allowed to fully relax by using a per atom force threshold of 10$^{-2}$ eV/bohr and a total energy threshold of 10$^{-3}$ eV. A scalar relativistic ultrasoft pseudopotential[48] and the Perdew−Burke−Ernzerhof (PBE) exchange-correlation functional[49] were used in this study, with a converged plane-wave energy cutoff of 680 eV. The equivalent converged $k$-space sampling density for the six-atom unit cell is 15 × 15 × 4 in the reciprocal-space Bi[$\overline{1}$10], Bi[$\overline{11}$2], and Bi[111] directions, respectively. These runtime parameters yielded a total energy convergence of 1 meV/atom. The atomic positions in the slab were fully relaxed for each of the simulations presented. The number of $k$-points was scaled to the slab dimensions, with 1 $k$-point used for both the out-of-plane reciprocal-space sampling direction (Bi[111]) and for the in-plane direction Bi[$\overline{1}$10] with 21 atoms in the unit cell, while 15 were used in the other in-plane direction Bi[$\overline{11}$2]. The in-plane slab dimension were selected as a result of the step width optimization discussed in the Results section.

## RESULTS

**Cleaved Surface of Bi(111).** The cleaved surface of Bi(111) consists of large atomically flat terraces with steps heights of ∼4.0 Å, corresponding to a single bilayer. The step edges are reasonably ordered (see Figure 1a), and the step edge directions reflect the 6-fold symmetry of the (111) plane. The well-ordered nature of the surface is demonstrated by STM and LEED measurements. Atomic resolution can be observed in Figure 1b,c. The line profile in Figure 1f illustrates that the interatomic distance is around 4.5 Å, which is in agreement with other works.[36,50] The LEED measurements presented in Figure 1d demonstrates that the cleaved crystal is highly ordered and has long-range order. XPS measurements (Figure 1e) demonstrate that this surface is atomically clean without measurable levels of carbon or oxygen contamination.

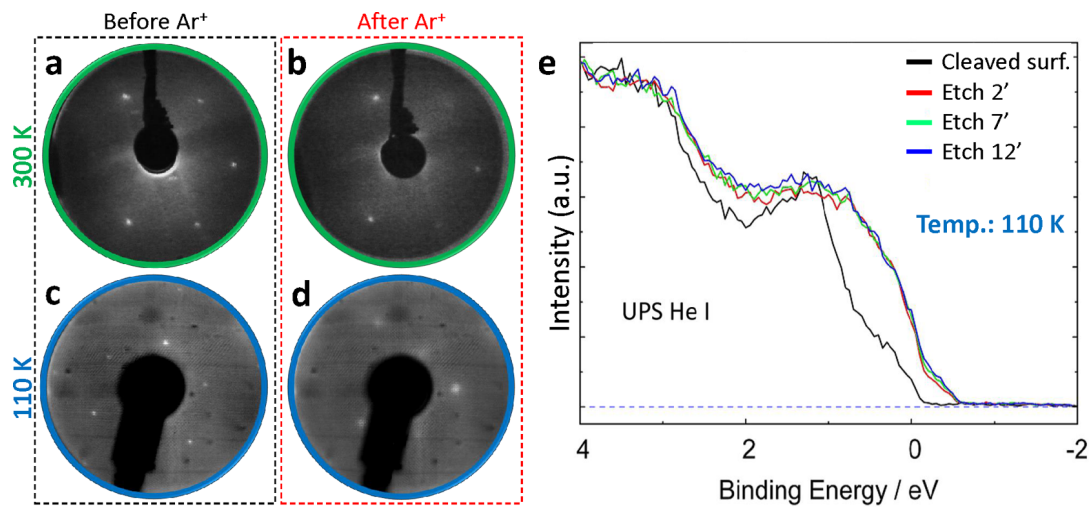

**Figure 3.** LEED and UPS spectra of the Bi(111) surface following Ar$^+$ etching. (a) LEED pattern was obtained from the cleaved surface of Bi(111) at room temperature at $E_p$ = 57 eV. (b) LEED after the 10 min of sputtering at room temperature and at $E_p$ = 57 eV ($E$ = 2 keV, $P_{Ar}$ = 5 × 10$^{-5}$ mbar, and $I$ = 20 μA). (c) LEED of the cleaved crystal at 110 K and at $E$ = 95 eV. (d) LEED of the sputtered surface at 110 K for 12 min ($E$ = 2 keV, $P_{Ar}$ = 5 × 10$^{-5}$ mbar, and $I$ = 20 μA). LEED in (c) and (d) obtained in conjunction with UPS measurements. (e) UPS measurements were performed at 110 K before and after different durations of Ar$^+$ etching ($E$ = 2 keV, $P_{Ar}$ = 5 × 10$^{-5}$ mbar, and $I$ = 20 μA). The black UPS spectra correspond to the UHV cleaved Bi(111) prior to Ar$^+$ etching. Ar$^+$ sputtering was conducted at 110 K for 2 min (red), 7 min (green), and 12 min (blue).

Following XPS measurements, UPS measurements were performed on the same cleaved surface, which can be seen in Figure 3e (black line).

The electronic structure of Bi type B step edges has been investigated. A large-scale STM image of Bi(111) terraces (150 × 150 nm$^2$) obtained at a bias of 1.2 V and a tunneling current of 70 pA is presented in Figure 2a. A smaller scale STM image (30 × 30 nm$^2$) corresponding to the area indicated by the green square in Figure 2a is presented in Figure 2b. Two terraces separated by terrace edge, which run vertically though the center of the image ($V$ = 1.0 V and $I$ = 80 pA). STS measurements were performed at each of the 50 points along the 15 nm length blue line.[43,51] An $I(V)$ curve was obtained for each point[52] and differentiated with respect to the voltage. Figure 2c displays representative d$I$/d$V$ spectra measured on and away from the step edge. (The blue triangle is taken from point 5, the red triangle is acquired from point 25, and the green triangle is taken from point 45 in Figure 2d.) The resulting d$I$/d$V$ curves are shown in Figure 2d as a 2D map. Each vertical line is a d$I$/d$V$ curve plotted between −0.5 and +0.5 V, with the magnitude of d$I$/d$V$ expressed by using a color scale, where warmer colors indicate a larger value. Spectra were measured every 3 Å along the blue line in Figure 2b. The d$I$/d$V$ spectrum at point number 25 was performed on the central step (Figure 2b). A suppression of the local density of states (LDOS) near the Fermi level (in the range of −0.01 to −0.25 V) is observed at the double bilayer step edges. We note the similarity between points 5 and 45, obtained on the each terrace, which highlights the quality of the STS measurement. The 2D map shows the evolution of the edge state is space; the blue region, corresponding to lower DOS, reduces as one moves away from the step. At points 5 and 45 (blue and green spectra in Figure 2c) the step is clearly not influencing the electronic characteristics. Calculated and experimental LDOS of the bilayer terraces are compared in Figure 1S of the Supporting Information. The bilayer calculation is in qualitative agreement with the corresponding experimental STS data with the minima slightly shifted to the right of the Fermi level.

**Surface of Bi(111) after Argon Bombardment.** 2D defects were introduced to the surface via argon sputtering, and the influence on the surface periodicity was examined by LEED. The changes to the electronic structure in the surface region were monitored by UPS measurements during the sputtering process. An argon ion beam energy of 2 keV was used at an incidence angle of 55° (emission current, $I$ = 20 μA). Bismuth samples were bombarded at 300 and 110 K for 10 and 12 min, respectively. With regards to the sputtering parameters, sample size, and design of our chamber the approximate flux, fluence, and dosage are 3 × 10$^{11}$ ions/cm$^2$ s, 2 × 10$^9$ ions/s, and 1.8 × 10$^{15}$ ions/cm$^2$, respectively.

LEED images taken before and after (Figures 3a and 3b, respectively) Ar$^+$ sputtering for 10 min at 300 K indicate that the surface remains ordered after sputtering. Evidently, the surface of the Bi crystal recrystallizes following the disordering induced by Ar$^+$ sputtering. In an effort to prevent the recrystallization process, the experiment was repeated at 110 K. By cooling the crystal, we aimed to retain the amorphous structure of the surface, as has been demonstrated in the case of Sb(111).[53] In that work, Chekmazov et al. did not observe a LEED pattern after Ar$^+$ bombardment at low temperature (110 K); however, a diffraction pattern appeared as the temperature was increased to 300 K. The presence of a LEED pattern following Ar$^+$ sputtering at for 12 min at 110 K indicates that the surface, once again, possesses sufficient energy to regain crystallinity. Sputtering at temperatures below 110 K, which have not been explored in this study, may be required to retain the amorphous state. The terminating bismuth layer can be amorphous or the lattice can be distorted, which can change the electronic structure of Bi(111).[54] However, taking into account the mean free path of electrons incident to Bi and the energy range employed, the LEED pattern is predominantly formed from a depth of ∼5 Å.[55] Spot profile analyses of the LEED patterns (at 110 and 300 K) before and after Ar$^+$ treatments are shown in Figure 2S.

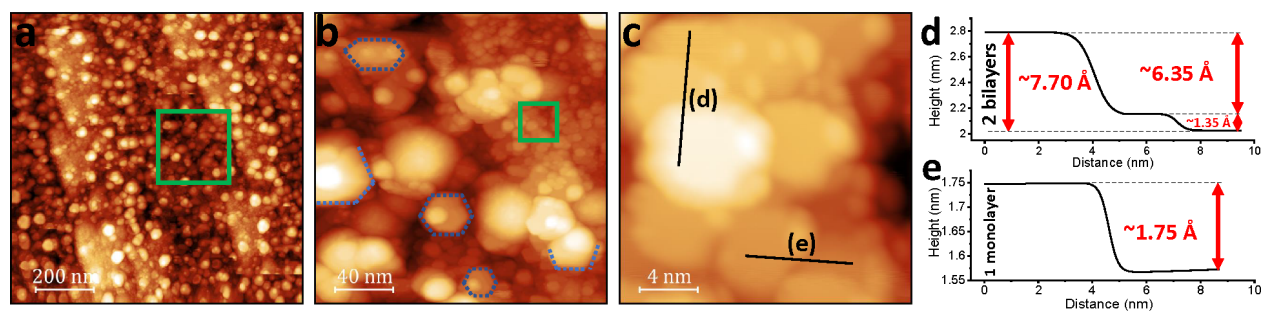

**Figure 4.** STM images after Ar$^+$ sputtering. Ar$^+$ sputtering at room temperature with ion beam energy of 2 keV and partial pressure $P_{Ar} = 5 \times 10^{-5}$ mbar for 10 min. Scale and the scanning parameters for (a) (1000 × 1000 nm$^2$, V = 1.5 V, and I = 73 pA), for (b) (200 × 200 nm$^2$, V = 1.5 V, and I = 70 pA), and for (c) (20 × 20 nm$^2$, V = 1.5 V, and I = 70 pA). (d) and (e) are line profiles in image (c) which demonstrate a step height of the Bi monolayer of around ~1.35 and ~1.75 Å.

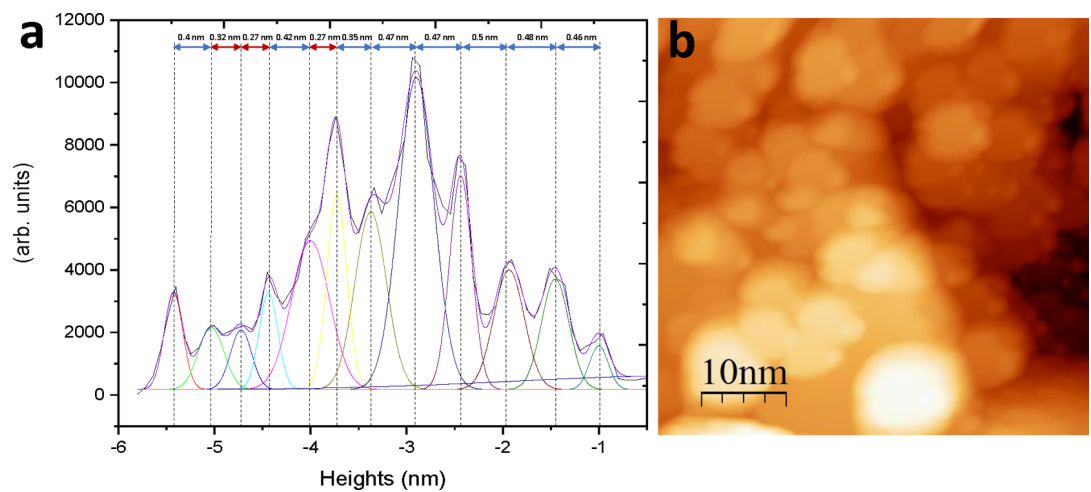

**Figure 5.** Histogram of step heights following Ar$^+$ sputtering. (a) Deconvolved step heights of the sputtered surface shown in (b); step heights corresponding to monolayer steps (≈2.7 Å) are indicted by red arrows. (b) STM image after Ar$^+$ sputtering at room temperature with partial pressure $P_{Ar} = 5 \times 10^{-5}$ mbar for 10 min. Scale and scanning parameters of the image are 50 × 50 nm$^2$, V = 1.5 V, and I = 80 pA.

UPS measurements indicate that the DOS near the Fermi level increases as the surface is bombarded. UPS measurements were performed before sputtering and after 2, 7, and 12 min of sputtering at 110 K. The corresponding spectra are presented in Figure 3e. It is apparent that the intensity of the UPS spectra near the Fermi level increases, approximately by a factor of 3, following 2 min of Ar$^+$ etching. Thus, following sputtering the surface exhibits more pronounced metallic properties. Further sputtering of the surface did not result in a remarkable increase in the UPS spectrum near the Fermi level. This observation is in contrast with the study of Sb(111),[53] which demonstrated that the feature near the Fermi level edge becomes more pronounced with increasing etching duration, corresponding to the more defective surface. In the case of Bi(111), one can conclude that 2 min is sufficient to saturate the surface with defects and that this saturation limit is related to the recrystallization process. To see a gradual shift of the Fermi edge for Bi, the sputtering conditions (time, ion energy, or flux) would need to be decreased.

STM measurements were performed following Ar$^+$ sputtering at 300 K. Figure 4a depicts a large-scale STM image (1000 × 1000 nm$^2$, V = 1.5 V, and I = 73 pA), which demonstrates the change in topography of the Bi(111) surface in comparison to the cleaved surface presented in Figure 1a. While the surface structure initially appears disordered, closer inspection reveals order on a smaller scale. This observation is verified by LEED measurements (Figure 3b). Figure 4b depicts a subsection of Figure 4a (200 × 200 nm$^2$, V = 1.5 V, and I = 70 pA) which displays pseudohexagonal structures (nanoislands) with preferential edges (indicated with dashed, blue lines) corresponding to the 6-fold symmetry of Bi(111). Drozdov et al.[8] have shown that hexagonal "pits" in Bi(111) exhibit alternating step edge types, A and B. Bi(111) islands are geometrically identical and demonstrate the same alternating step edge types. Sputtering results in a significant increase to the number of type A and type B bilayer edges at the surface.

The STM image (20 × 20 nm$^2$, V = 1.5 V, and I = 70 pA) presented in Figure 4c displays a close-up of these structures. Issues with tip stability, as a result of scanning this rough sputtered surface, limited the quality of STS measurements, and hence, STS measurements are not presented. The line profile across a step edge of the nanoisland shown in Figures 4d and 4e demonstrates a step heights of approximately 1.35 and 1.75 Å, closely corresponding to the monolayer step height of 1.6 Å.[9,34,56,57]

Further analysis of the step heights was conducted, and the resulting histogram is displayed in Figure 5a. The histogram shows the number of points (from Figure 5b) as a function of the tip−surface distance, relative to the highest terrace on the surface. To determine the step heights, each peak is deconvoluted and the distance between peaks measured. From the histogram presented in Figure 5a, the monolayer terraces are estimated to comprise ∼25% of the surface. Two distinct step heights are apparent, 4 and 2.7 Å. The 4 Å step corresponds to a bilayer of Bi(111) while the value of 2.7 Å is assigned to a monolayer of Bi(111). The theoretical atomic lattice spacing between single Bi(111) layers is 1.6 and 2.4 Å[34] depending on the bond type, covalent or van der Waals. The surface relaxation of the monolayer atoms and a distinct LDOS, demonstrated in the next section, affect the STM tip–surface distance, and thus the measured values differ slightly than that of theory. The smaller step height difference of 1.6 Å is not seen in this histogram; however, step heights of 1.35 and 1.75 Å are observed in Figure 4c and may correspond to this step. The peaks corresponding to these 1.35 and 1.75 Å might be masked by the other larger peaks. The small separation between this peak and the common bilayer peaks makes deconvolution challenging. The height profiles indicate the presence of an additional step feature, with step height values close to that of the theoretical lattice spacing for a monolayer of Bi(111).[34,35]

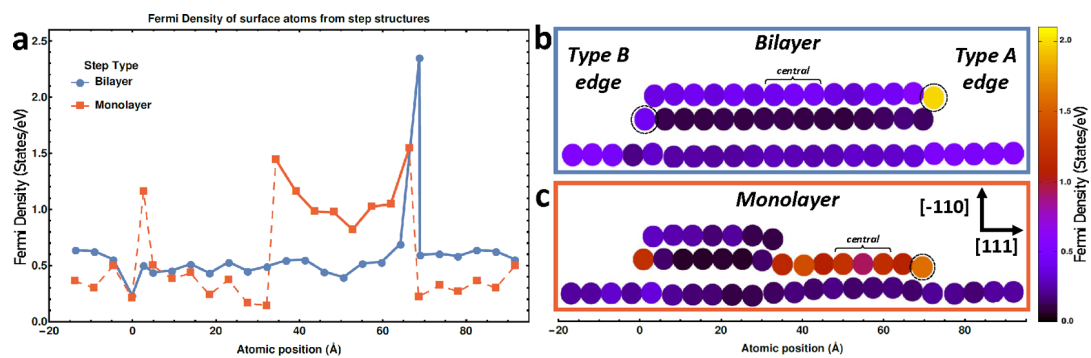

Figure 6. Simulated Fermi density (LDOS at the Fermi level) of monolayer and bilayer steps on Bi(111): (a) Fermi density of surface atoms from bilayer (blue) and monolayer (red) step structures ((b) and (c), respectively), calculated by using Löwdin population analysis.[61] The line for the monolayer section of the monolayer step structure is solid, while the remaining bilayer component is broken. (b) Atomic positions in the surface region for a bilayer step exhibiting a type A and type B step edge. The color of each filled circle indicates the magnitude of Fermi density. (c) Atomic positions in surface region for a monolayer step structure. The color of each filled circle indicates the magnitude of Fermi density. The nine simulated atomically thin layers under the surfaces shown in (b) and (c) are not depicted. In (a) note the large Fermi density of the A type step and the monolayer, in comparison with the bilayer and B type step, which more normally dominate the cleaved surface.

Considering that the energy of Ar$^+$ ions is on the order of 10$^3$ eV and that the binding energy (either covalent or van der Waals) is of the order of 1 eV, the probability of an ion breaking covalent or van der Waals bonds is approximately equal. This may then explain the presence of the monolayers structures on the surface. Sun et al.[58] demonstrated that covalent bonds on Bi(100) break following bombardment with Ar$^+$ bombardment with an energy of 300 eV. The presence of ordered structures following argon sputtering leads to the conclusion that two opposing processes are involved in their formation: amorphization and recrystallization of the surface. The sputtering process destroys the surface order, and the concurrent recrystallization is likely due to the diffusion of atoms on the surface. If so, it is a thermally activated process with a very low activation energy. At 110 K, the diffusion process results in the partial recrystallization of the surface and the presence of normally energetically unfavorable but stable in UHV monolayer steps. The increased density of bilayer steps will not increase the Fermi level character; STS measurements in Figure 2 show the bilayer step has a lower DOS around the Fermi level.

The present findings indicate that argon sputtering produces these 2D defects (monolayer steps and nanoislands), increasing the DOS near the Fermi level as indicated by UPS measurements in Figure 3e. Similarly to the case of Sb(111),[53] the formation of monolayer steps is a result of a local violation to the conditions for the Peierls transition. The Peierls transition is an out-of-plane distortion to the periodic lattice of Bi(111), which results in the formation of a layered structure with alternating covalent and van der Waals bonds. These bonds are energetically favorable in comparison to the simple cubic structure. Local breaking of these bonds results in an energetically unfavorable monolayer structure at the surface. This process leads to a change in the spectrum of electronic states: the proportion of monolayers (with higher DOS) will increase, which may lead to the evolution in the UPS spectra near the Fermi level in Figure 3e.

Recently, the breakage of covalent bonds of a layered material was observed in the GeTe(111) crystal.[59] Like Bi, GeTe along the [111] direction has short strong bonds (covalent) and long weaker bonds (van der Waals).[60] In the study of the GeTe(111) crystal, it has been shown that the Ge termination, which corresponds to covalent bonds, results from cleaving. Surface vacancies, which may originate from Ar$^+$ sputtering, may also play a role in increasing the Fermi density at the surface. Calculations by Sahoo et al. indicate an increase in the Fermi density of the atoms that surround the vacancy due to dangling bonds.[22] Any contribution that may arise from surface vacancies is not believed to significantly contribute to the Fermi density. Good LEED images after Ar$^+$ sputtering indicative of Bi(111) imply vacancies are at low concentration if present, whereas the histogram shows a significant number of monolayer features on the surface.

It should be noted that the crystal was not moved during cycles of Ar$^+$ sputtering and UPS measurements. Furthermore, the work function of the material does not change following ion etching. Additionally, XPS survey spectra following Ar$^+$ sputtering do not indicate the presence of Ar or any other elements on the surface of Bi. XPS spectra of the carbon 1s, oxygen 1s, and argon 2p regions obtained after Ar$^+$ sputtering are presented in Figure 3S.

**Density Functional Theory Calculations of Step Structures.** DFT calculations were performed to further investigate the structural and electronic properties of Bi(111) step types (monolayer and bilayer) and step edges (type A and type B). The calculations help elucidate the experimental measurements, which suggest that the monolayer terraces

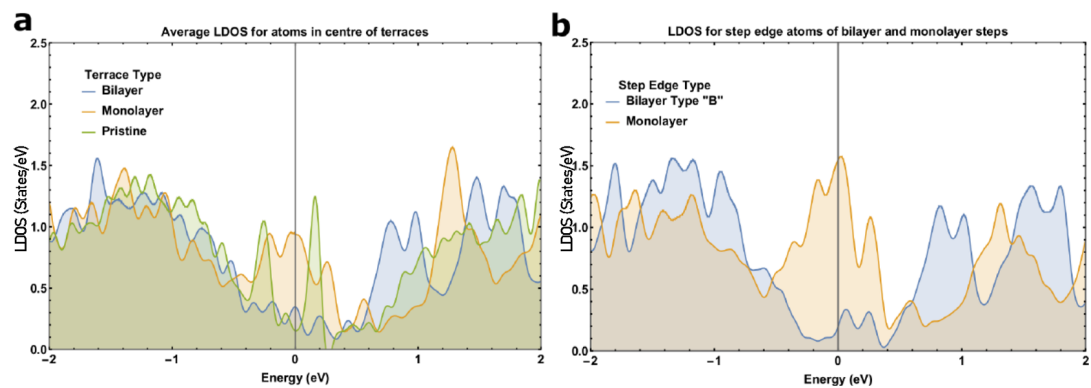

**Figure 7.** Simulated LDOS of atoms from monolayer and bilayer terraces of Bi(111) (a) Average LDOS of the central 3 atoms from the bilayer step of Figure 6b and the 3 central atoms of the monolayer step in Figure 6c. LDOS of the pristine surface of a 60-layer Bi(111) slab is also presented. (b) LDOS of the bilayer type B and monolayer step edge atoms indicated in Figures 6b and 6c.

exhibit a higher DOS near the Fermi level, leading to more pronounced metallic properties.

The electronic and structural properties of a slab that includes a bilayer step were first investigated. The slab was composed of 5 complete bilayers with an additional partial bilayer at the surface, resulting in a terrace edge. Each layer of the 5 complete bilayers is composed of 21 atoms which run in the Bi$[\bar{1}10]$ direction. The number of atoms in the partial bilayer was varied to determine the sufficiently large terrace width. Slabs are allowed to fully relax, and the differences in the bond lengths between the central atoms of the partial bilayers are compared. The average difference in bond length between steps that have 7 and 17 atoms in the Bi$[\bar{1}10]$ direction was ∼0.5%. Here, 7 and 17 atoms comprise the top half of the bilayer, while the bottom half is composed of 8 and 18 atoms in the Bi$[\bar{1}10]$ direction. The bilayer step with 7 atoms in the Bi$[\bar{1}10]$ direction was the minimum step width used in simulations. A slab with a bilayer terrace terminated by type A and type B step edges was fully relaxed (see Figure 6b).

The Fermi density (LDOS at the Fermi level) was calculated by using Löwdin population analysis[61] based on pseudoatomic orbitals. The locations of the circles in Figure 6b indicate the atomic positions of the surface region of the system in the Bi$[\bar{1}10]$ and Bi$[111]$ direction, while the color of the circles indicates the magnitude of the (Kohn–Sham system) Fermi density. The blue line in Figure 6a indicates the Fermi density of the surface atoms as a function of atomic position in the Bi$[\bar{1}10]$ direction. The Fermi density is approximately the same for each atom on the surface, reducing near the type B step edge and increasing significantly at the type A step edge. The STS data presented in Figure 2c also exhibit a lower DOS near the Fermi level for the bilayer step edge compared to the terrace. A terrace terminated by two type B step edges was found to be more energetically favorable (∼0.21 meV/atom) than that type A and type B terrace depicted in Figure 6b.

Figure 6c displays the atomic positions of the surface region of a monolayer step structure following relaxation. This structure is achieved by adding a bilayer that is incomplete and asymmetric: the top (terminating) atomic layer of the bilayer contains 7 atoms while the underlying atomic layer contains 17 atoms in the Bi$[\bar{1}10]$ direction. For comparison, the 9 complete layers beneath the surface contain 21 atoms each. The color of the filled circles once again represents the magnitude of the Fermi density. The solid red line in Figure 6a indicates the Fermi density of the monolayer step portion of this structure, while the broken red line indicates the remaining portion, consisting of a bilayer step and terrace. The increase in the Fermi density in the region of the monolayer step by a factor of 2−3 may be noted, with the largest increase occurring at the monolayer step edges. The atoms in this section of the layer and the atoms from the layer beneath have relaxed toward one another, and as such, the spacing closely resembles that of covalent bonds rather than the van der Waals bonds. The change in spacing extends to neighboring atoms and results in a small deviation in the Fermi density as compared to the conventional bilayer atoms. Integrated LDOS plots from the Fermi level to −1 eV and from the Fermi level to +1 eV are presented in the Figure 4S.

Figure 7a displays the atom-averaged DOS within a ±2 eV window centered on the Fermi level for the three central surface atoms (shown in Figure 6b,c) of the bilayer and monolayer steps. Two distinct DOS curves are evident, with the monolayer step (orange) having a larger DOS in the vicinity of the Fermi level, importantly between 0 and −0.6 eV, in agreement with the differences observed by UPS. Figure 7b displays the LDOS in a ±2 eV window centered on the Fermi level for the step edge atoms (indicated in Figure 6b,c) of each step type. The monolayer edge atom (orange) has a larger DOS near the Fermi level. The LDOS of the pristine surface (green) is included for comparison. The DFT simulations suggest the type A bilayer step and monolayer terraces can increase the Fermi density in comparison to the bilayer terrace and type B bilayer step. This indicates that these 2D defects can increase the Fermi density of the sputtered surface, as is observed in experiment.

We cannot distinguish between type A and type B bilayer steps from the histogram in Figure 5. However, by considering both the predicted contributions to the Fermi DOS from the monolayer, the type B step edge and the experimental observation of the increased Fermi DOS, one can gain insight into the coverage of the monolayer in comparison to the type B step edge. One assumes each Bi surface atom occupies the same unit volume, and each of these volumes exhibits the LDOS of that atom throughout. As previously mentioned, the monolayer terraces are estimated to comprise ∼25% of the surface from the histogram presented in Figure 5. From DFT calculations presented in Figure 7a,b, each central monolayer terrace atom contributes approximately +0.5 eV$^{-1}$ to the Fermi density relative to the bilayer terraces. The monolayer edge atoms contribute nearly +1 eV$^{-1}$. The type B bilayer edge atoms reduce the Fermi density by ∼0.2 eV$^{-1}$. Let us consider the case where the total contribution to the Fermi DOS from the type B bilayer edge atoms is opposite and equal to the total contribution from the monolayer terrace; there must be 2.5 times more type B bilayer edge atoms than monolayer terrace atoms. As we estimate around a quarter of the atoms are monolayer terrace atoms, we require around 2/3 of all the atoms to be bilayer type B edge atoms, for cancellation. This is not possible as edge atoms are greatly outweighed by terrace atoms. Furthermore, if one also considers the monolayer edge atoms, which contribute twice as much as the monolayer terrace atoms, the proportion of the atoms that need to be bilayer type B edge to achieve cancellation further increases. Finally, type A bilayer edge atoms will increase the Fermi density, and the histogram cannot distinguish between the type of bilayer terrace edge. Therefore, the monolayer contribution to the Fermi density outweighs that of the type B step edges, resulting in an increase to the Fermi density. The precise contributions to the Fermi DOS from states induced by sputtering are not clear; however, it is clear that these states, in some combination, increase the Fermi DOS.

## CONCLUSIONS

We have thoroughly investigated the nature of the physical and electronic structure of monolayer and bilayers Bi(111) steps and their corresponding step edges. Clean, atomically flat Bi(111) undergoes partial surface recrystallization following Ar$^+$ bombardment. This process occurs at temperatures as low as 110 K. One would assume that Ar$^+$ sputtering destroys the surface order, resulting in an amorphous state; however, simultaneous diffusion remarkably results in recrystallization and, consequently, nanostructures exhibiting monolayer steps. These structures have been observed by using STM measurements. LEED diffraction patterns, which are measured before and after ion bombardment, also indicates that the surface retains an ordered nature.

UPS measurements, taken at intervals during Ar$^+$ bombardment, indicate an increase in the number of electronic states near the Fermi level following bombardment. This increase in the DOS is understood to originate from two sources. First, the local breaking of the Peierls transition, which results in the presence of energetically unfavorable monolayer steps, contributes. DFT calculations predict the monolayer terraces exhibit a significantly larger DOS near the Fermi level than bilayer terraces. Second, the presence of hexagonal nanoislands on the surface increases the number of bilayer type A and type B step edges. DFT calculations predict the type A edges to exhibit a Fermi density even larger than that of the monolayer terraces. Conversely, type B edges exhibit a Fermi density smaller than that of the bilayer terraces. This indicates that the increase in the DOS observed by UPS is due to the increased presence of the monolayers and type A step edges. The Kohn−Sham density of states is qualitatively consistent with the zero-frequency limit of UPS, indicating that the Fermi density is boosted by a factor of 3 near the defected regions as a result of the breaking of the Peierls transition. The increased Fermi density at the surface observed in the monolayers is reminiscent of that of topological insulators, which see use in thermoelectrics and future applications in spintronics with topological quantum computations.[62,63]

## ASSOCIATED CONTENT

### *ⓈSupporting Information
The Supporting Information is available free of charge at https://pubs.acs.org/doi/10.1021/acs.jpcc.0c07345.

Calculated and experimental LDOS (Figure S1); spot profiles of the LEED measurements before and after Ar$^+$ bombardment (Figure S2); survey and zoomed-in XPS scans of the carbon 1s, oxygen 1s, and argon 2p regions after Ar$^+$ ion sputtering (Figure S3); integrated plots of calculated LDOS (Figure S4) (PDF)

## AUTHOR INFORMATION

### Corresponding Authors
**Kuanysh Zhussupbekov** − *School of Physics and Centre for Research on Adaptive Nanostructures and Nanodevices (CRANN), Trinity College Dublin, Dublin 2, Ireland*; orcid.org/0000-0003-1909-3270; Email: zhussupk@tcd.ie

**David D. O'Regan** − *School of Physics and Centre for Research on Adaptive Nanostructures and Nanodevices (CRANN), Trinity College Dublin, Dublin 2, Ireland; AMBER, the SFI Research Centre for Advanced Materials and BioEngineering Research, Dublin 2, Ireland*; orcid.org/0000-0002-7802-0322; Email: david.o.regan@tcd.ie

**Igor V. Shvets** − *School of Physics and Centre for Research on Adaptive Nanostructures and Nanodevices (CRANN), Trinity College Dublin, Dublin 2, Ireland*; Email: ivchvets@tcd.ie

### Authors
**Killian Walshe** − *School of Physics and Centre for Research on Adaptive Nanostructures and Nanodevices (CRANN), Trinity College Dublin, Dublin 2, Ireland*

**Brian Walls** − *School of Physics and Centre for Research on Adaptive Nanostructures and Nanodevices (CRANN), Trinity College Dublin, Dublin 2, Ireland*

**Andrei Ionov** − *Institute of Solid State Physics, Russian Academy of Sciences, Chernogolovka, Russia*

**Sergei I. Bozhko** − *School of Physics and Centre for Research on Adaptive Nanostructures and Nanodevices (CRANN), Trinity College Dublin, Dublin 2, Ireland; Institute of Solid State Physics, Russian Academy of Sciences, Chernogolovka, Russia*

**Andrei Ksenz** − *Institute of Solid State Physics, Russian Academy of Sciences, Chernogolovka, Russia*

**Rais N. Mozhchil** − *Institute of Solid State Physics, Russian Academy of Sciences, Chernogolovka, Russia*

**Ainur Zhussupbekova** − *School of Physics and Centre for Research on Adaptive Nanostructures and Nanodevices (CRANN), Trinity College Dublin, Dublin 2, Ireland*; orcid.org/0000-0003-2724-8762

**Karsten Fleischer** − *School of Physics and Centre for Research on Adaptive Nanostructures and Nanodevices (CRANN), Trinity College Dublin, Dublin 2, Ireland; School of Physical Sciences, Dublin City University, Dublin 9, Ireland*; orcid.org/0000-0002-7638-4480

**Samuel Berman** − *School of Physics and Centre for Research on Adaptive Nanostructures and Nanodevices (CRANN), Trinity College Dublin, Dublin 2, Ireland*; orcid.org/0000-0002-4321-387X

Ivan Zhilyaev − *Institute of Microelectronics Technology and High Purity Materials, Russian Academy of Sciences, Chernogolovka, Russia*

Complete contact information is available at:
https://pubs.acs.org/10.1021/acs.jpcc.0c07345

**Notes**
The authors declare no competing financial interest.

■ ACKNOWLEDGMENTS

This work was supported by SFI through the PI grant (12/IA/1264), Irish Research Council Laureate Award (IRCLA/2019/171), the AMBER Centre (12/RC/22782), the European Regional Development Fund (ERDF), the Government of the Republic of Kazakhstan under the Bolashak program, the Research Facility Center at the ISSP of RAS, Institute of Microelectrnics Technology and High Purity Materials, Russian Academy of Sciences, RFBR Grant 19-29-03021, and Erasmus Plus mobility grants (2016-1-IE02-KA107-000479, 2017-1-IE02-KA107-000538, and 2018-1-IE02-KA107-000589). This work was supported by TCHPC (Research IT, Trinity College Dublin). All calculations were performed on the Boyle cluster maintained by the Trinity Centre for High Performance Computing. This cluster was funded through grants from the European Research Council and Science Foundation Ireland.